\documentclass{aa}
\usepackage{psfig}
\def\tento#1{$\times$10$^{#1}$}

\def\SNII{{\rm SN}\,{\sc ii}}
\def\SNeII{{\rm SNe}\,{\sc ii}}
\def\HII{{\rm H}\,{\sc ii}}
\def\HI{{\rm H}\,{\sc i}}
\begin{document}

   \thesaurus{03     
              (11.09.1 NGC~4303;  
	       11.19.3;  
               11.01.2;  
	       13.25.2)} 

   \title{X-rays from the barred galaxy NGC 4303}


   \author{D. Tsch\"oke
          \inst{1}
          \and
          G. Hensler
          \inst{1}
	  \and
	  N. Junkes
	  \inst{2}
          }

   \offprints{D. Tsch\"oke}

   \institute{Institut f\"ur Theoretische Physik und Astrophysik, 
              Universit\"at Kiel, D-24098 Kiel, Germany\\
              email: tschoeke@astrophysik.uni-kiel.de
         \and
             Max-Planck-Institut f\"ur Radioastronomie, Auf dem 
             H\"ugel 69, D-53121 Bonn, Germany\\
             }

   \date{Received 20 March 2000 / Accepted 21 June 2000}

   \titlerunning{X-rays from the barred galaxy NGC~4303}
   \maketitle

   \begin{abstract}

The late-type galaxy NGC~4303 (M61) is one of the most intensively 
studied barred galaxies in the Virgo Cluster. Its prominent 
enhanced star formation throughout large areas of the disk can be 
nicely studied due to its low inclination of about 27\degr. 

We present observations of NGC~4303 with the ROSAT PSPC and HRI in 
the soft X-ray (0.1--2.4 keV). The bulk of the X-ray emission is 
located at the nuclear region. It contributes more than 80\% to 
the total observed soft X-ray flux. The extension of the central 
X-ray source and the $L_{\rm X}$/$L_{\rm H\alpha}$ ratio point to
a low luminous AGN (LINER) with a circumnuclear star-forming region.
Several separate disk sources can be 
distinguished with the HRI, coinciding spatially with some of the 
most luminous \HII\ regions outside the nucleus of NGC~4303. 
The total star formation rate amounts to 1--2 M$_{\sun}$ 
yr$^{-1}$. The X-ray structure follows the distribution of star 
formation with enhancement at the bar-typical patterns.

The best spectral fit consists of a power-law component 
(AGN and HMXBs) and a thermal plasma component of 
hot gas from supernova remnants and superbubbles.
The total 0.1--2.4 keV luminosity of NGC~4303 
amounts to 5\tento{40} erg s$^{-1}$, consistent with comparable 
galaxies, like e.g. NGC~4569.


   \end{abstract}


\section{Introduction}

Barred galaxies constitute a major fraction of all disc galaxies 
classified in the optical, more than 50\% including strong bars 
and intermediate morphologies (Sellwood \& Wilkinson 
\cite{sel93}). This fraction increases when also near-infrared 
images are used for classification, thus underlining the 
importance for the general understanding of the evolution of 
galaxies. The non-axisymmetric potential has a strong impact on 
the gas dynamics and the star formation in barred systems. 
Observations reveal a correlation between the radial abundance 
gradient and the strength of the bar (Martin \& Roy \cite{mar94}; 
Friedli et al. \cite{fri94}; Martinet \& Friedli \cite{mar97}). 
This is interpreted as the result of two effects caused by the bar: 
a stronger radial gas flow and hence a stronger radial mixing of 
metals and the efficiency of star formation. The radial mass 
transfer concentrates gas near the galactic center and at the ends 
of the bar at corotation. Enhanced star formation is the 
consequence of gas accumulation. The rotating bar potential also 
heats up the outer disk parts which leads to larger stellar 
velocity dispersions and a radial diffusion of stars. 
(Sellwood \& Wilkinson \cite{sel93}).

Galactic bars have also been considered to support the central 
infall of gas to feed a central "monster"(e.g. Beck et 
al.~\cite{bec99}). Several authors have 
claimed that active galactic nuclei (AGN) are more likely in 
barred galaxies than in non-barred ones (e.g. Simkin et al. 
\cite{sim80}; Arsenault \cite{ars89}). Hummel et al. 
(\cite{hum90}) note that the fraction of central radio sources in 
barred spirals is by a factor of 5 higher than in non-barred 
spirals. Other authors doubt that there is a significantly higher 
number of bars in galaxies harboring an AGN (e.g. Balick \& 
Heckman \cite{bal82}; Ho et al. \cite{ho97}). It appears that the 
concentration of gas on a scale of $\sim$1 kpc at the galactic 
center required to enhance the central star formation can easily
be achieved by a bar potential. It seems much more difficult, 
however, to accumulate enough gas on a scale of a few pc to tens 
of pc in order to produce an AGN . Other effects depending on 
the environment of the 
galaxies (interaction: Elmegreen et al. \cite{elm90}; \HI\ 
contents: Cayatte et al. \cite{cay90}) play an important role in 
mass distribution, gas flow, and therefore in the formation 
and evolution of bars and the star formation history in these 
systems.

One of the most famous, closest and most widely studied barred 
galaxies is NGC~4303 (M61), member of the Virgo Cluster, which 
is observed at an inclination of 27\degr\ (Guhathakurta et al. 
\cite{guh88}). Optical spectra of this galaxy indicate that it 
consists of a nuclear starburst and a LINER or Seyfert 2 nucleus 
(Filippenko \& Sargent \cite{fil86}; Kennicutt et al. 
\cite{ken89}; Colina et al. \cite{col97}; Colina \& Arribas 1999, 
hereafter \cite{col99}). Indications for a high star formation 
rate (SFR) in NGC~4303 are given by the numerous \HII\ regions 
(Hodge \& Kennicutt \cite{hod83}; Martin \& Roy 1992, hereafter 
\cite{mar92}) and three observed supernovae (van Dyk 
\cite{van92}). It also shows strong radio emission distributed 
over the entire disk (Condon \cite{con83}). Colina et al. 
(\cite{col97}) and \cite{col99} discussed the question of a 
starburst--AGN connection in this barred galaxy, using optical 
spectroscopy and HST UV images. The data range from a nuclear 
spiral structure of massive star-forming regions with an outer 
radius of 225 pc down to the unresolved core of a size $\le$ 
8 pc. From the UV data it is not clear if the core is a massive 
stellar cluster or a pure AGN. 

VLA observations (Cayatte et al. \cite{cay90}) show that NGC~4303 
is not highly \HI\ deficient, which can be explained by only slight 
environmental influences in the outermost region of the Virgo 
Cluster. The projected distance to M87 is 8\fdg2 (Warmels 
\cite{war88}). No significant difference of the abundance gradient 
in the disk of NGC~4303 compared to non-barred spiral galaxies has 
been observed (\cite{mar92}). Martinet \& Friedli (\cite{mar97}) 
discussed the abundance gradient slope in terms of bar age. 
According to them, a steep gradient in the bar and a flat one in 
the outer disk are typical for a young bar while a single flat 
gradient in bar and disk characterizes an old bar. 
\cite{mar92} did not determine the 
gradient at large radii because of a small number of \HII\ regions. 
Martinet \& Friedli (\cite{mar97}) also claimed that bars in 
late-type spirals with enhanced star formation like NGC~4303 are 
expected to be young. 

Probable interaction companions are the nearby galaxies 
NGC~4303\,A (Condon \cite{con83}) and NGC~4292 (Cayatte et al. 
\cite{cay90}), at distances of 7\farcm5 northwest and 10\arcmin\ 
northeast, respectively.

\begin{table}
\caption{Some basic parameters of NGC~4303.}
\begin{tabular}{lcc}
\hline
 & NGC 4303 & Reference$^a$ \\ \hline
alternative name & M61 & \\
 & IRAS 12194+0444 & \\
type & SAB(rs)bc & 1\\
RA (2000) & 12$^\mathrm{h}$21$^\mathrm{m}$54\fs9 & 2\\
Dec (2000) & +04\degr 28\arcmin 25\arcsec & 2\\
distance & 16.1 Mpc & 3\\
$D_{25}$ diameter & 5\farcm9 & 4\\
axis ratio & 0.97 & 4\\
inclination & 27\degr & 5\\
log $L_\mathrm{B}$ & 10.53$^b$ & 4\\
log $M_\mathrm{H}$ & 9.71$^b$ & 4\\
\hline
\label{tabgalpar}
\end{tabular}
\vspace{-2mm}
\\
$^a$References:\\
1) de Vaucouleurs et al. (\cite{dev91})\\
2) NASA/IPAC Extragalactic Database\\
3) Ferrarese et al. (\cite{fer96})\\
4) Tully (\cite{tul88})\\
5) Guharthakurta et al. (\cite{guh88})\\
$^b$Corrected for a distance of 16.1 Mpc.\\
\end{table}


\section{Observations and data reduction}

In this paper we present data from the High Resolution Imager 
(HRI) and the Position Sensitive Proportional Counter (PSPC) on 
board of the X-ray satellite ROSAT. This X-ray telescope was 
operating in the energy range of 0.1--2.4 keV. For details 
concerning ROSAT and its detectors see the ROSAT User's Handbook 
(Briel et al. \cite{bri96}).


\subsection{HRI}

We proposed HRI pointed observations of NGC~4303 (ROSAT sequence 
number 600854, PI: N. Junkes), which were carried out in the time 
periods July 2--8, 1996, June 9--23, 1997, and January 6--7, 1998 
with integration times of 16055 sec, 27538 sec, and 5600 sec, 
respectively (see Table~\ref{tabobs}). All three observations are 
centered on RA(2000) = 12$^\mathrm{h}$21$^\mathrm{m}$55\fs2; 
Dec(2000) = +04\degr28\arcmin11\farcs5. 

The three photon lists were combined to create one single image. 
The data were analyzed using the commercial interactive data 
language software package IDL (Interactive Data Language). 
All presented images contain pixel 
values with units of counts per pixel and second. The images were 
corrected for vignetting but not for background. To quantify any 
source counts three circular source-free fields with radii of 
49\arcsec, 34\arcsec, and 39\arcsec, respectively, were selected 
to determine the background level (Table~\ref{tabobs}). The 
averaged background flux amounts to 3.52\tento{-7} cts s$^{-1}$ 
arcsec$^{-2}$.

\begin{figure}
\psfig{figure=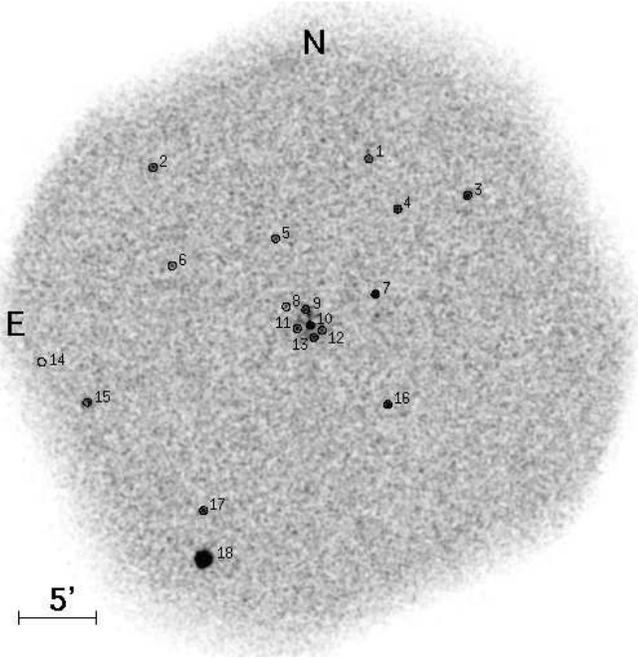,width=8.8cm}
\caption{Total field of view of the combined ROSAT HRI observation 
of NGC~4303. The detected 5$\sigma$ sources are labeled with 
numbers. The sources in the center (nos. 9--13) correspond to 
NGC~4303.}
\label{hrifov}
\end{figure}

Fig.~\ref{hrifov} shows the HRI field of view (FOV) of the three 
combined data sets, convolved with a Gaussian of 5\arcsec\ FWHM. 
For reasons of identification the 5$\sigma$ sources detected with 
a maximum likelihood method are numbered in Fig.~\ref{hrifov} and 
listed in Table~\ref{tabdetsub}. Several single source detections 
(no. 9--13) coincide with optical peaks in the galaxy. Source 
no.~18 can be identified as the QSO 1219+044, which served as the 
central source of the PSPC pointed observation used for spectral 
analysis in Sect.~\ref{pspcobs}. Source no.~7 coincides spatially 
with the QSO 1219+047 (Bowen et al. \cite{bow96}). It turned out 
that the source is variable the X-rays. Its flux increased 
by a factor of 3 between the second and third HRI observation 
(June 1997 and January 1998).

\begin{table*}
\caption{Date, integration time, and source and background count 
rates of the HRI and PSPC observations.}
\begin{tabular}{clccc}
\hline
 & Date & Integration time & Count rate & Background \\
 & & [s] & [cts s$^{-1}$] & [cts s$^{-1}$ arcsec$^{-2}$]\\ \hline
HRI & July 2--8, 1996 & 16055 & 8.8\tento{-3}\ $^a$ & 
3.52\tento{-7}\ $^a$\\
 & June 9--23, 1997 & 27538 & &\\
 & January 6--7, 1998 & 5600 & & \\
PSPC & December 24--26, 1992 & 8135 & 0.062 & 5.14\tento{-7}\\ 
\hline
\label{tabobs}
\end{tabular}
\vspace{-2mm}
\\
$^a$Combined HRI observations\\
\end{table*}


\begin{table*}
\caption{EXSAS source detection of X-ray sources in the HRI 
coinciding with the optical image of NGC~4303.}
\begin{tabular}{clccc}
\hline
Source no. & RA & Dec & Count rate & Identification\\
 & (2000) & (2000) & [10$^{-4}$ cts s$^{-1}$] \\ \hline
1 & 12$^\mathrm{h}$21$^\mathrm{m}$39\fs8 & 
+04\degr 39\arcmin 17\arcsec & 12.5$\pm$2.4 & \\
2 & 12$^\mathrm{h}$22$^\mathrm{m}$35\fs7 & 
+04\degr 38\arcmin 44\arcsec & 16.5$\pm$2.9 & \\
3 & 12$^\mathrm{h}$21$^\mathrm{m}$14\fs1 & 
+04\degr 36\arcmin 54\arcsec & 26.1$\pm$3.0 & GSC 00285-00416\\
4 & 12$^\mathrm{h}$21$^\mathrm{m}$32\fs3 & 
+04\degr 36\arcmin 01\arcsec & 13.5$\pm$2.1 & 
unidentified optical point source\\
5 & 12$^\mathrm{h}$22$^\mathrm{m}$03\fs9 & 
+04\degr 34\arcmin 06\arcsec & 6.4$\pm$1.4 & \\
6 & 12$^\mathrm{h}$22$^\mathrm{m}$30\fs7 & 
+04\degr 32\arcmin 19\arcsec & 7.9$\pm$1.7 & \\
7 & 12$^\mathrm{h}$21$^\mathrm{m}$38\fs0 & 
+04\degr 30\arcmin 28\arcsec & 47.2$\pm$3.2 & 
QSO 1219+047; variable\\
8 & 12$^\mathrm{h}$22$^\mathrm{m}$01\fs2 & 
+04\degr 29\arcmin 40\arcsec & 5.3$\pm$1.3 & \\
9 & 12$^\mathrm{h}$21$^\mathrm{m}$56\fs1 & 
+04\degr 29\arcmin 29\arcsec & 10.8$\pm$2.0 & NGC~4303\\
10 & 12$^\mathrm{h}$21$^\mathrm{m}$54\fs9 & 
+04\degr 28\arcmin 27\arcsec & 36.1$\pm$3.0 & NGC~4303\\
11 & 12$^\mathrm{h}$21$^\mathrm{m}$58\fs3 & 
+04\degr 28\arcmin 15\arcsec & 8.5$\pm$1.6 & NGC~4303\\
12 & 12$^\mathrm{h}$21$^\mathrm{m}$51\fs8 & 
+04\degr 28\arcmin 08\arcsec & 6.9$\pm$1.5 & NGC~4303\\
13 & 12$^\mathrm{h}$21$^\mathrm{m}$54\fs0 & 
+04\degr 27\arcmin 40\arcsec & 11.9$\pm$2.0 & NGC~4303\\
14 & 12$^\mathrm{h}$23$^\mathrm{m}$04\fs4 & 
+04\degr 26\arcmin 03\arcsec & 29.4$\pm$5.1 & \\
15 & 12$^\mathrm{h}$22$^\mathrm{m}$52\fs6 & 
+04\degr 23\arcmin 26\arcsec & 28.8$\pm$3.5 & \\
16 & 12$^\mathrm{h}$21$^\mathrm{m}$34\fs8 & 
+04\degr 23\arcmin 17\arcsec & 21.8$\pm$2.3 & \\
17 & 12$^\mathrm{h}$22$^\mathrm{m}$22\fs5 & 
+04\degr 16\arcmin 23\arcsec & 18.8$\pm$2.8 & \\
18 & 12$^\mathrm{h}$22$^\mathrm{m}$22\fs5 & 
+04\degr 13\arcmin 16\arcsec & 351.6$\pm$9.2 & QSO 1219+044\\ 
\hline
\label{tabdetsub}
\end{tabular}
\end{table*}
 

\subsection{PSPC}
\label{pspcobs}

\begin{figure}
\psfig{figure=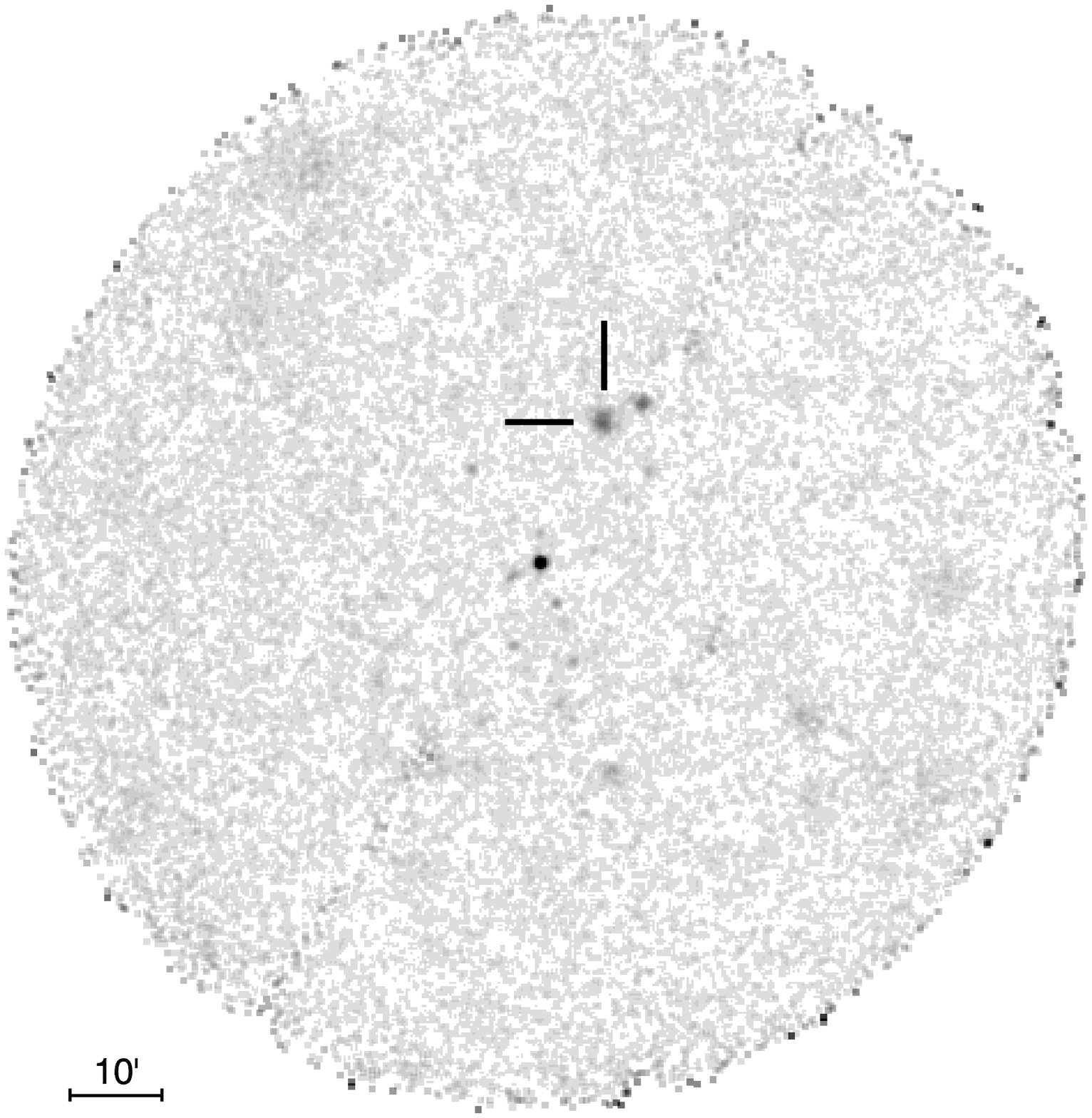,width=8.8cm}
\caption{Total field of view of the ROSAT PSPC observation around 
the QSO 1219+044. The position of NGC~4303 at an off-axis distance 
of 17\arcmin\ is marked.}
\label{pspcfov}
\end{figure}

Since there exist no particular PSPC observations pointed on 
NGC~4303, we use ROSAT observations from the archive (ROSAT 
sequence number 701095, PI: R. Staubert). 
NGC~4303 lies in the FOV at an 
off-axis distance of 17\arcmin\ which is still within the inner 
part of the supporting ring structure of the telescope. The 
position of the X-ray counterpart of NGC~4303 is given in 
Fig.~\ref{pspcfov}. The central source is the QSO 1219+044 
(no. 18 in Fig.~\ref{hrifov}). The image has been convolved with 
a Gaussian of 25\arcsec\ FWHM. Due to strong asymmetry and 
broadening of the point spread function (PSF) with radial distance 
from the optical axis ($\sim$25\arcsec\ FWHM for an on-axis point 
source at 1 keV; $\sim$67\arcsec\ FWHM for a 17\arcmin\ off-axis 
point source at 1 keV; Hasinger et al. \cite{has94}) we cannot get 
any useful spatial information from the PSPC data.

This PSPC exposure was carried out between December 24 and 26, 
1992, with an integration time of 8135 sec. The background flux 
was determined from three circular source-free areas in the field 
with radii of 83\arcsec, 88\arcsec, and 75\arcsec\ and amounts to 
5.14\tento{-7} cts s$^{-1}$ arcsec$^{-2}$. To analyze the PSPC 
data we used IDL and the software package XSPEC (Arnaud 
\cite{arn96}) for interactively fitting X-ray spectra.

For NGC~4303, we adopt the same distance as of M100, the brightest 
spiral in the Virgo Cluster (16.1 Mpc; Ferrarese et al. 
\cite{fer96}). Then the resolution of the ROSAT detectors of 
$\sim$5\arcsec\ full width at half maximum (FWHM) for the HRI and 
$\sim$67\arcsec\ FWHM for the PSPC at an off-axis distance of 
17\arcmin\ scale to 390 pc and 5.3 kpc, respectively.


\section{Results}


\subsection{Spatial analysis}

\begin{figure}
\psfig{figure=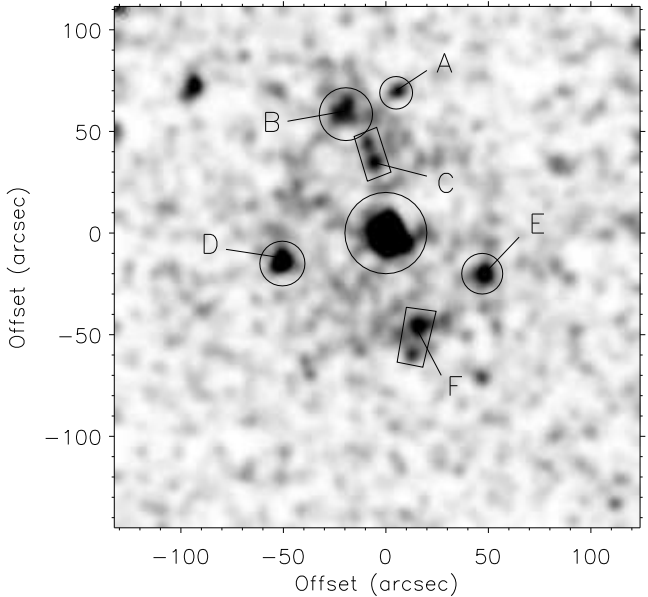,width=8.8cm}
\caption{The inner 4\arcmin\ of the pointed HRI observation. The 
labeled sources (A--F) are the ones spatially coinciding with the 
optical disk of NGC~4303 (see text and Table~\ref{tabhrisources} for 
details). The image is centered on the coordinates of the pointed 
observation. The distances on the axes are relative to the central 
X-ray source with the coordinates $\alpha$(2000) = 
12$^\mathrm{h}$21$^\mathrm{m}$55\fs5; $\delta$(2000) = 
+04\degr28\arcmin28\farcs5. North is up, east is to the left.}
\label{hrisrcareas}
\end{figure}


\begin{figure*}
\psfig{figure=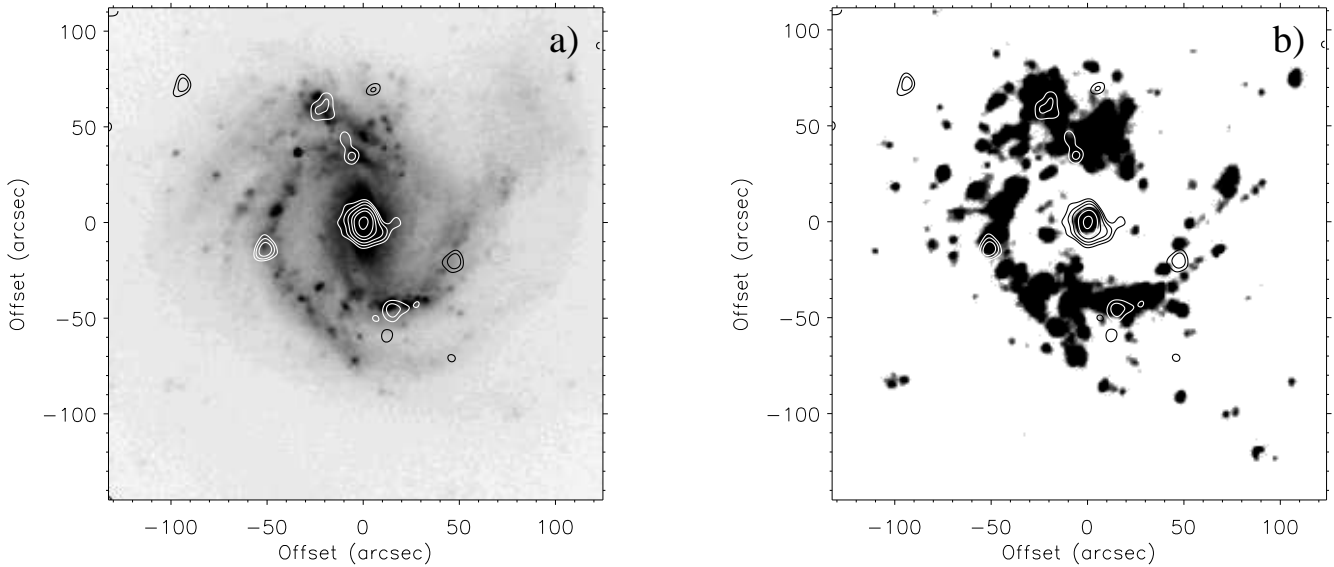,width=18cm,angle=-90}
\caption{Overlay of the HRI X-ray contours onto the optical image 
of NGC~4303 ({\bf a}) taken from Frei et al. (\cite{fre96}) and 
the H$\alpha$ image ({\bf b}) taken from \cite{mar92}. The contour 
levels are at 5, 7, 10, 15, 20, and 30$\sigma$ above the mean 
background with $\sigma$ = 1.5\tento{-7} cts s$^{-1}$ 
arcsec$^{-2}$. The images are centered on the coordinates of the 
pointed observation. The axis scales are relative to the central 
X-ray maximum at coordinates $\alpha$ = 
12$^\mathrm{h}$21$^\mathrm{m}$55\fs5; $\delta$ = 
+04\degr28\arcmin28\farcs5. North is up, east is to the left.}
\label{hrioveropt}
\end{figure*}


\begin{table*}
\caption{X-ray sources spatially coinciding with the optical 
galaxy NGC~4303. The source designations refer to 
Fig.~\ref{hrisrcareas}.}
\begin{tabular}{ccccccc}
\hline
Source & RA & Dec & Count rate & Source area & Flux$^a$ (0.1--2.4 keV) 
& Luminosity (0.1--2.4 keV)\\
 & (2000) & (2000) & [10$^{-4}$ cts s$^{-1}$] & [arcsec$^2$] & 
[10$^{-14}$ erg s$^{-1}$ cm$^2$] & [10$^{38}$ erg s${-1}$] \\ \hline
A & 12$^\mathrm{h}$21$^\mathrm{m}$54\fs5 & 
+04\degr 29\arcmin 38\arcsec & 3.4$\pm$1.1 & 200 & 1.4$\pm$0.5 & 
4.3$\pm$1.5 \\
B & 12$^\mathrm{h}$21$^\mathrm{m}$56\fs3 & 
+04\degr 29\arcmin 28\arcsec & 9.9$\pm$1.8 & 531 & 4.1$\pm$0.8 & 
12.7$\pm$2.5 \\
C & 12$^\mathrm{h}$21$^\mathrm{m}$55\fs3 & 
+04\degr 29\arcmin 03\arcsec & 5.9$\pm$1.4 & 290 & 2.5$\pm$0.6 & 
7.8$\pm$1.9 \\
D & 12$^\mathrm{h}$21$^\mathrm{m}$58\fs3 & 
+04\degr 28\arcmin 14\arcsec & 8.9$\pm$1.7 & 380 & 3.7$\pm$0.7 & 
11.5$\pm$2.2 \\
E & 12$^\mathrm{h}$21$^\mathrm{m}$51\fs7 & 
+04\degr 28\arcmin 08\arcsec & 7.5$\pm$1.5 & 315 & 3.1$\pm$0.6 & 
9.6$\pm$1.9 \\
F & 12$^\mathrm{h}$21$^\mathrm{m}$53\fs9 & 
+04\degr 27\arcmin 43\arcsec & 8.7$\pm$1.7 & 383 & 3.6$\pm$0.7 & 
11.2$\pm$2.2 \\
nucleus & 12$^\mathrm{h}$21$^\mathrm{m}$55\fs5 & 
+04\degr 28\arcmin 28\arcsec & 43.9$\pm$3.5 & 1321 & 87.8$\pm$7.0 & 
272$\pm$21 \\
\hline
\label{tabhrisources}
\end{tabular}
\vspace{-2mm}
\\
$^a$Energy conversion factor ECF:\\
\hspace*{1mm} disk sources A--F, ECF=2.4\tento{10} cts cm$^2$ erg$^{-1}$; 
nucleus, ECF=5\tento{13} cts cm$^2$ erg$^{-1}$\\
\end{table*}


As already mentioned, no useful spatial information on the X-ray 
structure of NGC~4303 can be obtained from the PSPC data due to the 
spatial resolution of the detector, and moreover to the 17\arcmin\ 
off-axis position of the source in the FOV.

In contrast, the more detailed HRI image reveals a number of X-ray 
sources distributed over the galactic disk 
(Fig.~\ref{hrioveropt}). In comparison to the numbered sources in 
Fig.~\ref{hrifov} the closer view allows to distinguish more 
details. For example, source no.~9 in Fig.~\ref{hrifov} splits 
into three X-ray spots (labeled A--C in Fig.~\ref{hrisrcareas}). 
The most luminous source coincides with the center of NGC~4303 and 
dominates in the soft X-rays. The count rates and fluxes derived 
for sources from the HRI are listed in Table~\ref{tabhrisources}. The 
corresponding areas are plotted in Fig.~\ref{hrisrcareas}. To 
determine the fluxes we used the energy conversion factor (ECF) 
from the ROSAT {\it Call For Proposals} documentation. The ECF 
determines the ratio between count rates and unabsorbed source 
flux in the ROSAT band for given spectral parameters. For the disk 
sources A--F we assume a 0.3 keV Raymond-Smith model  (Raymond \& 
Smith \cite{ray77}) with an absorption column density of 
3\tento{20} cm$^{-2}$. For the nucleus a power law with 
$\Gamma$=2.6 and a column density of 3\tento{20} cm$^{-2}$ (see 
Sect.~\ref{specfit} and Table~\ref{fittab}) is applied as spectral 
model.

The contours of sources B, C, D, and F in Fig.~\ref{hrisrcareas} 
are all located within the optical arm structure and coincide with 
bright H$\alpha$ emission regions within the spiral arms 
(Fig.~\ref{hrioveropt}). In addition, source E is embedded in the 
faint outer part of the southwestern spiral arm. \cite{mar92} 
distinguished 79 \HII\ regions in NGC~4303, mainly in the spiral 
arms. The X-ray contour overlay over the H$\alpha$ image in 
Fig.~\ref{hrioveropt} reveals that the X-ray sources B, C, D, and 
F coincide with \HII\ regions, while sources A and E are located 
near such regions.

Gas dynamical models of barred galaxies (Englmaier \& Gerhard 
\cite{eng97}) show strong gas accumulation at the tips of the bars 
due to corotation of the bar structure with the disk what should 
lead to enhanced star formation. \HI\ observations as well as the 
existence of prominent 
H$\alpha$ features strikingly support the outcome of these 
models. The X-ray contours B and F seem to arise from 
these regions. The X-ray maximum D is connected with another 
interesting feature of NGC~4303: in the eastern part the 
galactic arm seems to be deformed to a boomerang-like bow where 
source D lies at the bend but without any significant brightening 
in H$\alpha$.

The lower X-ray contours of the nucleus indicate a possible 
extended source. Recent high-resolution UV observations of the 
central region with the Hubble Space Telescope reveal a 
spiral-shaped structure of massive young (2--3 Myr) star-forming 
regions with an outer radius of 225 pc (\cite{col99}). This 
structure cannot be resolved by HRI. Due to the low age of the 
star clusters almost no thermal X-ray emission is expected at the 
galactic nucleus (see Sect.~\ref{discussion}). The extended X-ray 
contours may originate from additional sources at distances of 
about 1 kpc around the nucleus.

No X-ray emission has been detected from the possible interaction 
companions NGC~4303~A and NGC~4292. 


\subsection{Properties of the X-ray spectrum}
\label{specfit}

\begin{figure}
\psfig{figure=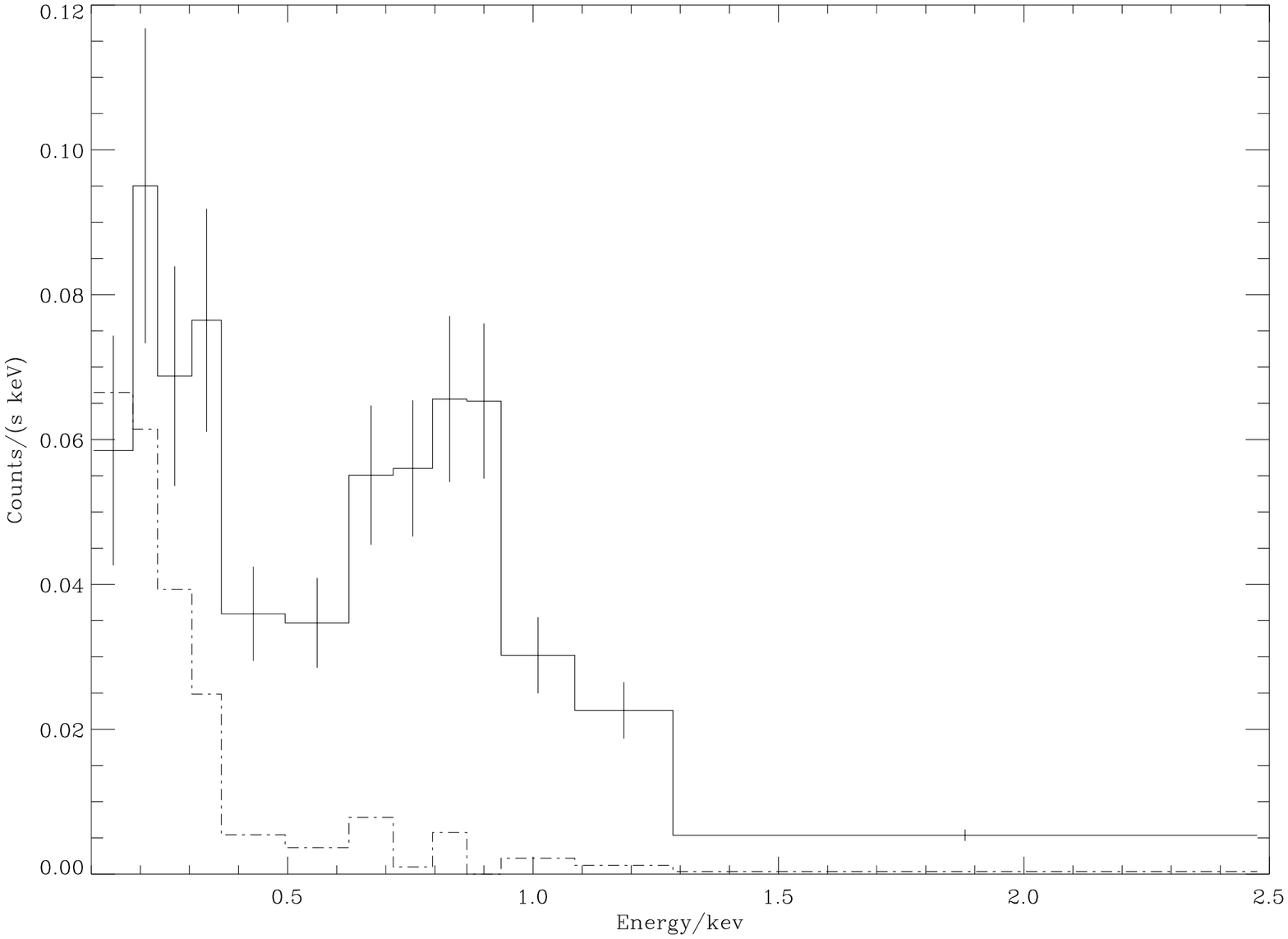,width=8.8cm,angle=90}
\caption{Background-subtracted ROSAT PSPC spectrum of NGC~4303 in 
the energy range of 0.1--2.4 keV. The 505 photons were binned to 
get a signal-to-noise ratio of 6. The dashed-dotted line 
represents the contribution of the background (see text).}
\label{pspcspec}
\end{figure}

Since the HRI maxima are separated by only 50\arcsec, and 
because of the reasons mentioned in Sect.~\ref{pspcobs}, the PSPC 
observations do not allow to study the spectra of the X-ray 
components of NGC~4303 individually. ROSAT PSPC detected 
505$\pm$24 backgroung-subtracted source counts from NGC~4303 in a 
total integration time of 8135 sec. The spectra of the source and 
the background are shown in Fig.~\ref{pspcspec}.


We fitted the spectrum with several single-component models, as 
Bremsstrahlung (BS), Raymond-Smith model (RS), and a power law 
(PO), and a combined RS-PO model. The results are listed in 
Table~\ref{fittab}.


\begin{table*}
\caption{Best spectral fits to the PSPC data of NGC~4303 and the 
derived model parameters. For the determination of the X-ray 
luminosity a distance of 16.1 Mpc is assumed.}
\begin{tabular}{llllllllll} 
\hline
Model & $N_\mathrm{H}$ & $kT$ & $\Gamma$ & $Z$ & Norm & 
Red. $\chi^2$ & d.o.f. & $F_\mathrm{X}$ & $L_\mathrm{X}$\\
(1) & (2) & (3) & (4) & (5) & (6) & (7) & (8) & (9) & (10)\\ 
[1.5mm]
\hline\\
RS + PO & 3.34$^{+1.95}_{-0.39}$ & 0.32$^{+0.14}_{-0.07}$ & 
2.6$^{+0.2}_{-0.6}$ & 1 & 6.2 (RS) & 1.3 & 8 & 
1.50$^{+0.48}_{-0.44}$ & 4.7$^{+1.5}_{-1.4}$ \\
 & & & & & 1.5 (PO) & & & (0.19$^{+0.04}_{-0.01}$) & 
(0.59$^{+0.12}_{-0.03}$)\\ [2mm]
PO & 5.72$^{+0.68}_{-0.50}$ & & 3.2$\pm$0.2 & & 2.1 & 1.4 & 10 & 
4.18$^{+1.93}_{-0.88}$ & 12.9$^{+6.0}_{-2.7}$ \\ [2mm] 
RS & 2.87$^{+0.67}_{-0.53}$ & 0.57$^{+0.08}_{-0.09}$ & & 0.006 & 
1.4 & 1.4 & 9 & 1.13$^{+0.08}_{-0.10}$ & 3.47$^{+0.25}_{-0.31}$\\ 
[2mm] 
BS & 3.19$^{+0.53}_{-0.49}$ & 0.56$^{+0.09}_{-0.07}$ & & & 2.5 
& 1.4 & 10 & 1.26$^{+0.12}_{-0.10}$ & 3.91$^{+0.37}_{-0.31}$\\[1mm]
\hline
\label{fittab}
\end{tabular}
\vspace{-2mm}
\\
Col. (1)--- Spectral models: BS = thermal Bremsstrahlung, RS = 
Raymond-Smith, PO = power law.\\ 
Col. (2)--- Column density in units of 10$^{20}$ cm$^{-2}$.\\ 
Col. (3)--- Plasma temperature in units of keV.\\ 
Col. (4)--- Photon index.\\ 
Col. (5)--- Metallicity in units of Z$_{\sun}$.\\
Col. (6)--- Scaling factor: for BS in units of 
(10$^{-18}$/(4$\pi D^2$))$\int n_\mathrm{e}n_\mathrm{I}$d$V$, 
$n_\mathrm{e}, n_\mathrm{I}$ = electron and ion densities 
(cm$^{-3}$); for RS in units of 
(10$^{-19}$/(4$\pi D^2$))$\int n_\mathrm{e}n_\mathrm{H}$d$V$, 
$n_\mathrm{e}, n_\mathrm{H}$ = electron and H densities 
(cm$^{-3}$); for PO in units of 10$^{-4}$ photons~keV$^{-1}$ 
cm$^{-2}$ s$^{-1}$ at 1 keV.\\ 
Col. (7)--- Reduced $\chi^2$.\\ 
Col. (8)--- Degrees of freedom.\\ 
Col. (9)--- Unabsorbed X-ray flux in units of 10$^{-12}$ erg cm$^{-2}$ 
s$^{-1}$. Values in brackets give the contribution of the thermal 
component.\\ 
Col. (10)--- X-ray luminosity in units of 10$^{40}$ erg s$^{-1}$. 
Values in brackets give the contribution of the thermal component.
\end{table*}


A single power-law model implies the assumption, that the active 
nucleus of NGC~4303 dominates the X-ray emission. Furthermore, the 
sources detected by the HRI in the galactic disk would also have to 
be described with the same power law. The photon index in this model 
is $\Gamma$=3.2$\pm$0.2. The emission of an AGN in the ROSAT 
energy band is best described by a power law with a photon index 
of $\Gamma\sim$2.4; nevertheless some cases have been observed 
with $\Gamma>$3 (MCG --5-23-16: Mulchaey et al. \cite{mul93}; 
Mkn~335: Turner et al. \cite{tur93}). High-mass X-ray binaries 
(HMXB) found in young star-forming regions in the spiral arms have 
a similar spectral shape in the 0.1--2.4 keV energy range with a 
photon index of $\Gamma\sim$2.7 (Mavromatakis \cite{mav93}). The 
column density of the absorbing component amounts to 5.7\tento{20} 
cm$^{-2}$, which is by a factor of 3 higher than the Galactic 
foreground \HI\ column density (Dickey \& Lockman 1990, 
\cite{dic90}). Nevertheless, self-absorption within NGC~4303 must 
be expected, and small-scale deviations from the observed Galactic 
value by \cite{dic90} cannot be ruled out and may result in a higher 
absorption from the Milky Way. The resulting 0.1--2.4 keV X-ray 
luminosity amounts to 1.3\tento{41} erg s$^{-1}$. The flux portion 
from the sources outside the nuclear region as observed with the 
HRI amounts to $\sim$1.4\tento{40} erg s$^{-1}$ in the case of a 
single power-law emission model with $\Gamma$=3.2 using the 
corresponding ECF of 1\tento{10} cts cm$^2$ erg$^{-1}$. Assuming a 
mean X-ray luminosity of 10$^{37}$ erg s$^{-1}$ for an HMXB, as 
observed in the Milky Way (Fabbiano et al. \cite{fab82}; Watson 
\cite{wat90}) would require an unlikely high number of 1400 of 
these systems to produce the observed X-ray flux. The ratio of OB 
stars to HMXBs is assumed to be $\sim$500 (Fabbiano et al. 
\cite{fab82}). This means that a total number of  7\tento{5} OB 
stars would be required to account for the HMXB X-ray flux in 
NGC~4303. Even if we consider to have 10$^{5}$ OB stars in 
NGC~4303, as observed e.g. in Mkn~297 (Benvenuti et al. 
\cite{ben79}), it is still a factor of 7 higher than expected. 
Moreover, this is the required number only for the disk 
sources and would involve almost 1.5\tento{7} M$_{\sun}$ in 
massive stars with a Salpeter IMF and, by this, would require 
a moderately high SFR of about 15 M$_{\sun}$ yr$^{-1}$ in 
the disk. On the other hand, the corresponding supernova 
type\,{\sc ii} (\SNII) rate (0.1 yr$^{-1}$) should contribute 
to the X-ray emission via hot gas.

The single component models BS and RS show similar results. 
Consequently, we only achieve an adequate fit of RS with very low 
metallicity, e.g. the portion of emission lines to the spectrum is 
very small. In contrast, it is expected that emission lines of 
highly ionized elements, like Fe and Mg, should play an important 
role in the X-ray spectrum of \SNeII\ in starburst regions because 
of the nucleosynthesis of massive \SNII\ progenitor stars (Woosley 
\& Weaver \cite{woo85}). In both models the column density is 
about 3\tento{20} cm$^{-2}$ and the plasma temperature is 0.6 keV. 
For the BS model we get a total X-ray luminosity of 4\tento{40} 
erg s$^{-1}$, for the RS model it is 3.5\tento{40} erg 
s$^{-1}$. RS models with different higher metallicities yield 
unacceptable fits.

\begin{figure}
\psfig{figure=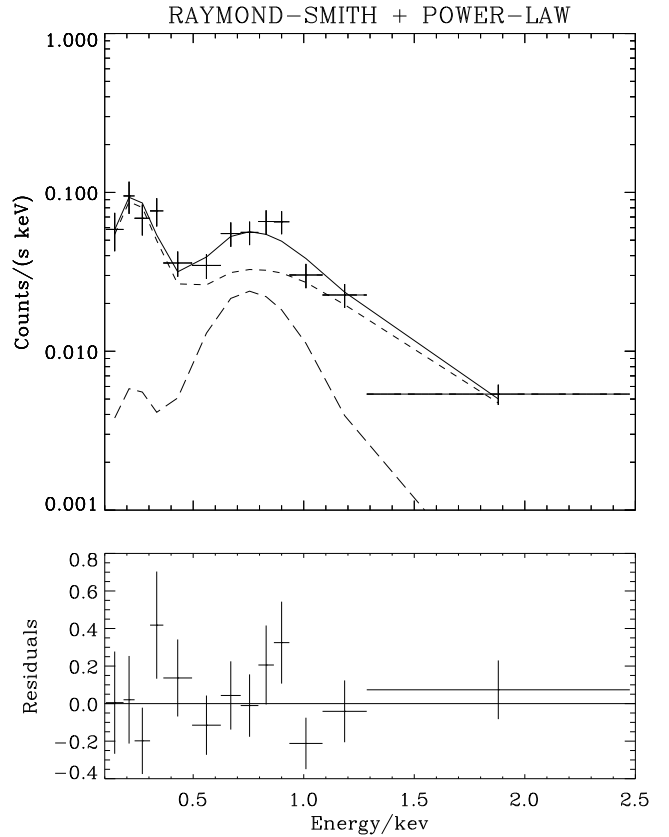,width=8.8cm}
\caption{Fit of the observed ROSAT PSPC spectrum of NGC~4303 with 
the two-component model RS+PO (solid line). The parameter values 
for this model are listed in Table~\ref{fittab}. The long-dashed 
line shows the spectrum of the thermal model component, the 
short-dashed line represents the power-law component. The 
residuals of the fit are plotted in the lower box.}
\label{rspomodel}
\end{figure}

The fit of the X-ray spectrum with a two-component model (RS+PO) 
is only slightly better than the one-component fits. Nevertheless, 
from the points mentioned above and the physical picture discussed 
in Sect.~\ref{discussion} this model serves as the best 
explanation for the observed soft X-ray emission. Hydrogen column 
density ($N_\mathrm{H}$=3.3\tento{20} cm$^{-2}$) and power-law 
spectral index ($\Gamma$=2.6) lie within the expected range (as 
discussed for the single power law above). The plasma temperature 
of 0.3 keV fits with the observed values of other galaxies (e.g. 
NGC~253: Forbes et al. \cite{for99}; NGC~1808: Junkes et al. 
\cite{jun95}). The total 0.1--2.4 keV luminosity for this model 
amounts to 4.7\tento{40} erg s$^{-1}$ with 13\% contribution 
from the RS component. The spectral fit together with the residuals 
is plotted in Fig.~\ref{rspomodel}.

\section{Discussion}
\label{discussion}

From the quality of the spectral fits alone there is no 
significance for favoring a single-component model or a 
combination of two components. Due to the lack of any spatial 
information in the PSPC image there is no possibility to 
distinguish between different spatial and spectral components 
simultaneously. So only the combined information from the PSPC and 
HRI data allows a more detailed interpretation of the X-ray 
results.

Several points speak against the single PO component model, as 
discussed in Sect.~\ref{specfit}. A more likely scenario is a 
composition of several different emission sources, like an active 
nucleus, HMXBs, and supernova remnants (SNRs). In the following 
we will therefore discuss a composite emission model and the 
comparison with the UV and optical observations of the galactic core.


\subsection{The nucleus of NGC 4303}

As can be discerned from the HRI image (Fig.~\ref{hrisrcareas} and 
Table~\ref{tabhrisources}), most of the X-ray emission of NGC~4303 
(83\%) comes from the central region of the galaxy. Three 
different pictures are imaginable for the nucleus: a central 
active nucleus, a central or circumnuclear region with enhanced 
star formation, or a combination of both phenomena. Any of these 
cases requires a sufficient gas density at the galactic center. 
This can be achieved by a barred potential which triggers radial 
gas flow from the outer regions toward the nucleus. On the other 
hand, from numerical simulations including gas dynamics bar 
formation has proved to be only a transient 
phenomenon (Combes \cite{com00}). In 
this picture, the galactic bar will be destroyed by the gas inflow 
after only a few cycles. A new bar-phase can follow this gas 
infall due to a subsequent gravitational instability from the 
accreted central mass. The problem with this picture is the 
contradiction of a necessary gas inflow to form and feed any 
nuclear activity (starburst and/or AGN) and the fact that this gas 
inflow destroys the bar. It seems that a sufficiently 
massive black hole can provide for its fuelling (Fukuda et al.
\cite{fuk98}). Another efficient way for gas to flow further into 
the center is a second smaller bar embedded into the first one 
due to a second inner Lindblad resonance (Friedli \& Martinet 
\cite{fri93}). In some cases, a gaseous circumnuclear ring is 
formed at the end of the second bar.

The concentrated X-ray emission from the galactic core in NGC~4303 
may originate from an AGN and/or a nuclear or 
circumnuclear starburst. A starburst can contribute in two 
different ways to the X-ray flux. First, the produced star 
population contains HMXBs, emitting an X-ray radiation in spectral 
shape similar to an AGN. HMXBs cannot be distinguished from AGN in 
the ROSAT data. Additionally, high-mass stars (above $\sim$8--10 
M$_{\sun}$) evolve to \SNeII\ at an age of $\sim$10$^7$ yr, 
depending on their initial mass. The \SNeII\ from one star cluster 
form a cumulative expanding superbubble filled with hot gas 
which can be described by a thermal Bremsstrahlung spectrum and 
additional emission lines and recombination edges of highly ionized 
heavy elements produced in high-mass stars and released by their 
\SNII\ explosions, e.g. O, Ne, Mg, Si, and Fe. Theoretical models 
for the spectral emission of such a hot diffuse gas are MEKA 
(Mewe et al. \cite{mew85}) and the model by Raymond \& Smith 
(\cite{ray77}), which we used in our spectral fit.

If we apply a two-component model to describe the X-ray spectrum 
of NGC~4303, the comparison of the flux ratio from the nucleus 
(83\% in the HRI) and the disk sources (17\% in the HRI) suggests 
that the central source can be described by the power-law 
component (87\% in the spectral fit). The thermal RS emission 
exclusively originates from the disk sources, indicating ongoing 
star formation. To consider the possible extension of 
$\sim$25\arcsec\ of the central source, as represented by the lowest 
contours, it is imaginable that a small fraction of the X-ray flux 
is emitted by a circumnuclear starburst at a distance of $\sim$1 kpc 
around the core. This would add a thermal component to the 
non-thermal X-ray nucleus. On the other hand, a fraction of the 
X-ray flux from the disk sources may come from HMXBs within these 
star forming regions.

Fuelling an AGN on scales of a few parsec at the center of the 
galaxy leads to the problem of reducing the angular momentum of 
the central gas by several orders of magnitude, as dynamical 
simulations show (Barnes \& Hernquist \cite{bar91}). Concentration 
of gas in a ring-like feature around the nucleus with a radius of 
$\sim$1 kpc is dynamically much easier to achieve. The HRI image 
agrees with the picture of an extended X-ray source with a 
diameter of the order of $\sim$2 kpc at the galactic center of 
NGC~4303, explained by a circumnuclear starburst region with an 
additional possible compact nuclear source.The decovered massive 
rotating circumnuclear disk in NGC~4303 can provide 
by its spiral-like structure of massive star forming regions an 
effective mechanism to channel gas from the circumnuclear regions 
further down to the nucleus to feed the AGN. But one has to 
consider that the spiral structure in the UV has a diamater of 
only 225 pc, while the extension of the central X-ray source is 
about 2 kpc in diameter. The spiral feature detected in the UV 
cannot be resolved with the HRI.

From the analysis of UV and optical magnitudes and colors of the 
central 250 pc Colina \& Wada (\cite{col00}) estimated ages of 
5--25 Myr for the star-forming regions. Consequently a contribution
to the central X-ray flux from SNRs and cumulatively expanding hot
gas has to be expected. The question remains whether we observe a pure 
nucleus of massive star-forming clusters or a composition of these
star clusters and a low luminous AGN. If NGC~4303 contains a 
non-thermal active nucleus, the X-ray luminosity of 4\tento{40} 
erg s$^{-1}$ points to only a low luminous AGN (LINER). 
Koratkar et al. (\cite{kor95}) found a correlation between $L_{\rm X}$
and $L_{\rm H\alpha}$ for low luminous AGNs of 
$L_{\rm X}$/$L_{\rm H\alpha} \approx$ 14. P\'erez-Olea \& Colina
(\cite{per96}) investigated the correlation between optical and X-ray
luminosities of several AGNs with circumnuclear star-forming rings,
pure AGNs, and pure starburst galaxies. The pure starbursts in their 
galaxy sample show $L_{\rm X}$/$L_{\rm H\alpha}$ values of 0.03--0.3, 
100 times smaller than for pure AGNs. If we take the H$\alpha$
luminosity of NGC~4303 derived by Keel (\cite{kee83}) and assume that
10\% originate from the nucleus, we get log $L_{\rm H\alpha}$(nucleus)
= 39.2 (adopted for a distance of 16.1 Mpc). Therefore the 
X-ray-to-H$\alpha$ ratio amounts to log($L_{\rm X}$/$L_{\rm H\alpha}$)
= 1.4, which agrees with the value found by Koratkar et al. 
(\cite{kor95}). Even the lower $L_{\rm X}$ value from a single RS model
($\sim$2.5\tento{40} erg s$^{-1}$ for the nucleus) results in 
log($L_{\rm X}$/$L_{\rm H\alpha}$) = 1.2. Typical pure starburst 
galaxies show H$\alpha$ luminosities of the order of their X-ray
luminosities or higher.


\subsection{The galactic disk}

At first glance the optical disk of NGC~4303 seems to have the 
quite symmetrical morphology of a late-type spiral. A closer look 
reveals that the eastern spiral arm of the galaxy has a much more 
prominent form with a boomerang-like shape and a lot more bright 
emission regions than the western counterpart. The northern disk 
shows a complex structure with many separate features. This 
asymmetry is more discernible in the H$\alpha$ image. The \HII\ 
regions are mainly located in the northern part of the disk at the 
junction of the bar with the eastern spiral arm and along that 
arm. A close encounter of one or both of the nearby galaxies 
NGC~4303\,A and NGC~4292 may have caused these features. The 
interaction within the Virgo Cluster is another possible source. 
Infall into the intracluster medium could cause ram pressure 
effects. Nevertheless, NGC~4303 is located at the outer edge of 
the cluster, which may produce only a moderate disturbance. This 
agrees with the \HI\ distribution over the whole optical disk. 
Galaxies lying nearer to the cluster center show \HI\ deficiencies 
and concentration of the neutral hydrogen in the central regions, 
indicating past interactions with the \HI\ gas been stripped off 
from the outer disk regions.

As a striking indication for accumulation of gas in these regions, 
the X-ray sources A--C and F within the galactic disk of NGC~4303 
are located at the ends of the bar. Gas dynamical simulations of 
barred galaxies have shown this accumulation due to mass flows 
along the bar to the center and to the ends of the bar, 
respectively (Noguchi \cite{nog88}; Englmaier \& Gerhard 
\cite{eng97}). The increased densities lead to enhanced star 
formation. 

From the low inclination of NGC~4303 no direct information can be 
obtained whether the disk is warped or not. But it is striking 
that source D lies exactly at the bend of the eastern 
boomerang-shaped arm. This may indicate that this X-ray source is 
caused by the tidal force leading to a local gas concentration. 
Another indirect hint for a past interaction comes from the 
spectra of the QSO 1219+047 (source no.~7 in Fig.~\ref{hrifov}), a 
QSO whose line-of-sight penetrates the outer \HI\ disk of 
NGC~4303. Bowen et al. (\cite{bow96}) detected complex Mg\,{\sc ii} 
absorption, spanning a velocity range of $\sim$300 km s$^{-1}$, 
despite the low inclined galactic disk. This high velocity is not 
fully understood. One possible explanation could be the result of 
interactions between NGC~4303 and the nearby companions.

\subsection{Star formation in the disk}

Table~\ref{tabhrisources} lists the observed count rates and 
derived 0.1--2.4 keV luminosities for the single X-ray sources in 
the disk of NGC~4303. These include the assumptions of a 
Raymond-Smith plasma with solar abundances at a temperature 
of 0.3 keV. A power-law model would increase the values by a factor 
of 2.5. A rough estimation of the SFR in the disk from the X-ray 
luminosities is done by calculating the \SNII\ rate 
$\nu_\mathrm{SN}$ using a SNR 
model by Cioffi (\cite{cio90}), and assuming a Salpeter IMF 
within a mass interval from 0.1 M$_{\sun}$ to 100 
M$_{\sun}$ and with all stars with masses above 8 M$_{\sun}$ 
evolving to \SNeII. According to Cioffi, a SNR expanding into an 
ISM with a density of 1 cm$^{-3}$ radiates a total energy of 
4.7\tento{49} erg in the soft X-ray regime above 0.1 keV for a time 
of $\sim$10$^4$ yr. Norman \& Ikeuchi (\cite{nor89}) investigated 
the cumulative effect of a number of SNRs. The total SFR 
in the disk of NGC~4303 from the 
X-ray luminosity amounts to 0.5 M$_{\sun}$ yr$^{-1}$. 
This however is just a very simple estimation, containing 
several simplifications, as e.g. the sum of the single SN model 
from Cioffi (\cite{cio90}) for several cumulatively expanding SNRs in
an evolving OB assoziation, or the derivation of the total disk SFR 
from single X-ray sources.
A more detailed determiation of the SFR, for example using an analytic
suberbubble model by Suchkov et a. (\cite{suc94}), would need 
information about the extensions and expansion time of the superbubbles, 
in order to determine the mechanical energy release by the SNe and, 
by this, the SN rate. This can be compared with the observed X-ray 
luminosity.

Besides the mentioned restrictions the difference between the SFR 
estimated from the X-ray flux and the 
SFR derived by the H$\alpha$ flux by Kennicutt (\cite{ken83}) 
(14 M$_{\sun}$ yr$^{-1}$) may be due to several reasons. Kennicutt 
used a Miller-Scalo IMF which increases the SFR by a factor of about 
1.5. The total H$\alpha$ luminosity underlies a distance 
determination with a Hubble constant of 50 km s$^{-1}$ Mpc$^{-1}$. 
Taking a radial velocity of 1569 km s$^{-1}$ (de Vaucouleurs et al. 
\cite{dev91}) leads to a 3.8 times higher luminosity than 
taking the distance of 16.1 Mpc which we used. Additionally, the 
fact that not all \HII\ regions may emit an adequate X-ray flux 
or are not strong enough to be detected lowers the estimated 
SFR from the X-ray 
luminosity which implies the existence of high-mass stars having 
been evolved to \SNeII. Possible X-ray emission from diffuse hot 
gas within the disk may lie below the detection limit of 1.1\tento{-6} 
cts s$^{-1}$ arcsec$^{-2}$ (5$\sigma$ above background level). 
A very faint component located at the spiral arms can be seen in 
outlines in Fig.~\ref{hrisrcareas}, but is not detected at a 3$\sigma$
level.
This limit corresponds to an X-ray flux of 4.6\tento{-17} erg s$^{-1}$ 
cm$^{-2}$ arcsec$^{-2}$ (ECF for a RS model as in 
Table~\ref{tabhrisources}). Kennicutt  (\cite{ken83})also 
admitted to treat the derived H$\alpha$ flux and resulting SFR 
with extreme caution because of only moderate accuracy due to 
possibly strong extinction effects. The H$\alpha$ flux derived by 
Keel (\cite{kee83}) is by a factor of 30 lower than the one derived 
by Kennicutt, after adopting the same distance.

Strikingly, the sources B and F both 
coincide with some of the most H$\alpha$ luminous \HII\ regions 
(Sources no.s 27 and 69 with log $F_{\mathrm H\alpha}$=$-$12.12 and 
log $F_{\mathrm H\alpha}$=$-$11.99, respectively, in \cite{mar92}). 
Depending on the fraction of the central X-ray flux steming from 
SNRs and superbubbles or from an AGN component, the SFR for the core 
is of the order of 1 M$_{\sun}$ yr$^{-1}$.


\section{Conclusions}

We have analyzed spatial and spectral data from the barred 
late-type spiral galaxy NGC~4303 in the soft X-ray regime. Several 
separate X-ray sources can be observed in the core and disk of the 
galaxy. The locations of the sources correspond to several \HII\ 
regions and indicate a concentration of gas at the center and at 
the ends of the galactic bar, in agreement with numerical 
simulations of gas dynamics in a barred potential. 

The low spatial resolution of the PSPC observation of NGC~4303 
does not allow a distinction of several individual X-ray sources 
within the object. The best fit of the soft X-ray spectrum taking 
into account the information from the high resolution HRI 
observation is a combination of a RS component with a temperature 
of 0.3 keV and a power-law component with a spectral index of 2.6. 
The total 0.1--2.4 keV X-ray luminosity amounts to 4.7\tento{40} 
erg s$^{-1}$, in agreement with other comparable barred galaxies 
with a nuclear starburst, like e.g. NGC~4569 (Tsch\"oke et al. in 
preparation). A pure starburst model for the nucleus of 
NGC~4303 would require a special explanation for the unusually 
high $L_{\rm X}$/$L_{\rm H\alpha}$ ratio.

The combination of the flux fraction of the separate sources, 
the spectral information, and the comparison with the H$\alpha$ 
luminosity from the core leads to the following picture: the soft 
X-ray emission originates from a composition of several distinct 
emission regions. The central source consists of a low luminous
AGN and a circumnuclear starburst. The disk sources are dominated by SNRs 
and superbubbles in star forming regions preferably at the ends of the 
bar and along the eastern spiral arm. Several HMXBs may contribute 
to the X-ray flux. 

The disk X-ray sources are coincident with some of the most luminous 
\HII\ regions in the galaxy. The estimated total SFR from the 
X-ray flux is 1--2 M$_{\sun}$ yr$^{-1}$. Most \HII\ regions are not 
detectable in the X-ray, like most H$\alpha$ sources in 
the eastern boomerang-shaped arm. The enhanced star formation 
in NGC~4303 may have been caused by some kind of interaction 
although the \HI\ morphology of the galaxy does not support 
very strong perturbation. If a dwarf galaxy has fallen 
in and merged with NGC~4303 in the past, the bar may have been 
produced with the subsequent triggering of the star formation at 
the center and in the spiral arms. The accreted dwarf galaxy would 
be resolved and not directly detectable.


\begin{acknowledgements}
The authors are grateful to Dominik Bomans for stimulating discussions, 
and to Dr. Olga Sil'chenko for her substantial and constructive report.
The ROSAT project is supported by the German Bundesministerium 
f\"ur Bildung, Wissenschaft, Forschung und Technologie (BMBF) and 
the Max-Planck-Society. This research has made use of the 
NASA/IPAC Extragalactic Database (NED) which is operated by the 
Jet Propulsion Laboratory, Caltech, under contract with the NASA. 
Observations made with the NASA/ESA Hubble Space Telescope were 
used, obtained from data archive at STScI. STScI is operated by 
the Association of Universities for Research in Astronomy, Inc. 
(AURA) under the NASA contract NAS 5-26555.
\end{acknowledgements}



\begin{thebibliography}{}
\bibitem[1996]{arn96}
Arnaud K.A., 1996, Astronomical Data Analysis Software and Systems 
V, ASP Conf. Ser. vol. 101, eds. Jacoby G., Barnes J., p. 17
\bibitem[1989]{ars89}
Arsenault R., 1989, A\&A 217, 66
\bibitem[1982]{bal82}
Balick B., Heckman T.M., 1982, ARA\&A 20, 431
\bibitem[1991]{bar91}
Barnes J.E., Hernquist L.E., 1991, ApJ 370, L65
\bibitem[1999]{bec99}
Beck R., Ehle M., Shoutenkov V., Shukurov A., Sokoloff D., 1999, 
Nat 397, 324
\bibitem[1979]{ben79}
Benvenuti P., Casini C., Heidmann J., 1979, Nat 282, 272
\bibitem [1996]{bow96}
Bowen D.V., Blades J.C., Pettini M., 1996, ApJ 472, L77
\bibitem [1996]{bri96}
Briel U., Aschenbach B., Hasinger G., et al., 1996, ROSAT User's 
Handbook (MPE, Garching)
\bibitem[1990]{cay90}
Cayatte V., van Gorkom J.H., Balkowski C., Kotanyi C., 1990, AJ 
100, 604
\bibitem[1990]{cio90}
Cioffi D., 1990, in "Physical Processes in Hot Plasmas", NATO ASI 
Series C, Vol. 305, eds. Brinkmann W., Fabian A.C., Giovanelli F., 
Kluwer, Dordrecht, p. 1
\bibitem[CA99]{col99} 
Colina L., Arribas S., 1999, ApJ 514, 637 (CA99)
\bibitem[1997]{col97}
Colina L., Garc\'{\i}a-Vargas M.L., Mas-Hesse J.M., Alberdi A., 
Krabbe A., 1997, ApJ 484, L41
\bibitem[2000]{col00}
Colina L., Wada K., 2000, ApJ 529, 845
\bibitem[2000]{com00}
Combes F., 2000, in "Galaxy Dynamics: from the Early Universe to 
the Present", ASP Conference Series, Vol. 197, eds. Combes F., 
Mamon G.A., Charmandaris V., p. 15
\bibitem[1983]{con83}
Condon J.J., 1983, ApJS 53, 459
\bibitem[1991]{dev91}
de Vaucouleurs G., de Vaucouleurs A., Corwin Jr. H.G., et al., 
1991, Third Reference Catalogue of Bright Galaxies, Springer 
Verlag, New York
\bibitem[DL90]{dic90}
Dickey J.M., Lockman F.J., 1990, ARA\&A 28, 215 (DL90)
\bibitem[1990]{elm90}
Elmegreen D.M., Elmegreen B.G., Bellin A.D., 1990, ApJ 364, 415
\bibitem[1997]{eng97}
Englmaier P., Gerhard O., 1997, MNRAS 287, 57
\bibitem[1982]{fab82}
Fabbiano G., Feigelson E., Zamorani G., 1982, ApJ 256, 397
\bibitem[1996]{fer96}
Ferrarese L., Freedman W.L., Hill R.J., et al., 1996, ApJ 464, 568
\bibitem[1986]{fil86}
Filippenko A.V., Sargent W.L., 1986, in "Structure and Evolution 
of Active Galactic Nuclei", eds. Giuricin G., Mardirossian F., 
Mezetti M., Ramella M., Dordrecht: Reidel, p. 21
\bibitem[1999]{for99}
Forbes D.A., Polehampton E., Stevens I.R., Brodie J.P., Ward M.J., 
1999, submitted to MNRAS, {\it astro-ph/9908048}
\bibitem[1996]{fre96}
Frei Z., Guhathakurta P., Gunn J.E., Tyson J.A., 1996, AJ 111, 174
\bibitem[1993]{fri93}
Friedli D., Martinet L., 1993, A\&A 277, 27
\bibitem[1994]{fri94}
Friedli D., Benz W., Kennicutt R., 1994, ApJ 430, L105
\bibitem[1998]{fuk98}
Fukuda H., Wada K., Habe A., 1998, MNRAS 295, 463
\bibitem[1988]{guh88}
Guhathakurta P., van Gorkom J.H., Kotanyi C.G., Balkowski C., 
1988, AJ 96, 851
\bibitem[1994]{has94}
Hasinger G., Boese G., Predehl P., et al., 1994, Legacy 4, 40, 
MPE/OGIP Calibration Memo CAL/ROS/93-015, Version: 1995, May 8
\bibitem[1997]{ho97}
Ho L.C., Filippenko A.V., Sargent W.L.W., 1997, ApJ 487, 591
\bibitem[1983]{hod83}
Hodge P.W., Kennicutt R.C., 1983, AJ 88, 296
\bibitem[1990]{hum90}
Hummel E., van der Hulst J.M., Keel W.C., 1990, A\&A 236, 333
\bibitem[1995]{jun95}
Junkes N., Zinnecker H., Hensler G., Dahlem M., Pietsch W., 1995, 
A\&A 294, 8
\bibitem[1983]{kee83}
Keel W.C., 1983, ApJS 52, 229
\bibitem[1983]{ken83}
Kennicutt R.C. 1983, ApJ 272, 54
\bibitem[1989]{ken89}
Kennicutt R.C., Keel W.C., Blaha C.A., 1989, AJ 97, 1022
\bibitem[1995]{kor95}
Koratkar A., Deustua S.E., Heckman T., et al., 1995, ApJ 440, 132
\bibitem[MR92]{mar92}
Martin P., Roy J.-R., 1992, ApJ 397, 463 (MR92)
\bibitem[1994]{mar94}
Martin P., Roy J.-R., 1994, ApJ 424, 599
\bibitem[1997]{mar97}
Martinet L., Friedli D., 1997, A\&A 323, 363
\bibitem[1993]{mav93}
Mavromatakis F., 1993, A\&A 273, 147
\bibitem[1985]{mew85}
Mewe R., Gronenschild E.H.B.M., van den Oord G.H.J., 1985, A\&AS 
62, 197
\bibitem[1993]{mul93}
Mulchaey J.S., Colbert E., Wilson A.S., et al., 1993, ApJ 414, 144
\bibitem[1988]{nog88}
Noguchi M., 1988, A\&A 203, 259
\bibitem[1989]{nor89}
Norman C.A., Ikeuchi S., 1989, ApJ 345, 372
\bibitem[1996]{per96}
P\'erez-Olea D.E., Colina L. 1996, ApJ 468, 191
\bibitem[1977]{ray77}
Raymond J.C., Smith B.W., 1977, ApJS 35, 419
\bibitem[1993]{sel93}
Sellwood J.A., Wilkinson A., 1993, Rep. Prog. Phys. 56, 173
\bibitem[1980]{sim80}
Simkin S.M., Su H.J., Schwarz M.P., 1980, ApJ 237, 404
\bibitem[1994]{suc94}
Suchkov A.A., Balsara D.S., Heckman T.M., Leitherer C., 1994, 
ApJ 430, 511
\bibitem[1988]{tul88}
Tully R.B., 1988, Nearby Galaxies Catalog, Cambridge University 
Press, Cambridge
\bibitem[1993]{tur93}
Turner T.J., George I.M., Mushotzky R.F., 1993, ApJ 412, 72
\bibitem[1992]{van92}
van Dyk S.D., 1992, AJ 103, 1788
\bibitem[1988]{war88}
Warmels R.H., 1988, A\&AS, 72, 427
\bibitem[1990]{wat90}
Watson M.G., 1990, in "Windows on Galaxies", eds. Fabbiano G., 
Gallagher J.S., Renzini A., Kluwer, Dordrecht, p. 177
\bibitem[1985]{woo85}
Woosley S.E., Weaver T.A., 1985, in "Radiation Hydrodynamics in 
Stars and Compact Objects", Proc. of the IAU Colloquium No. 89, 
eds. Mihalas D., Winkler K.-H., p. 91
\end{thebibliography}
\end{document}